\documentstyle[12pt,aasms4,epsf]{article}
\received{00  1999}
\accepted{00 1999}
\centerline{Submitted to the Editor of the ApJ Letters 
 on February 4, 1999}
\lefthead{Yusef-Zadeh and Cotton}
\righthead{OH (Radio Continuum  Observations}

\def\msol   {\hbox{$M_\odot$}}                  
\def\dsec   {\hbox{$.\!\!^{\rm s}$}}            

\begin{document}
\title{The  Position of Sgr A$^*$ at the Galactic Center}

\author{F. Yusef-Zadeh and D. Choate}
\affil{Department of Physics and Astronomy, Northwestern University, 
Evanston, Il. 60208 (zadeh@nwu.edu)}

\author{W. Cotton}
\affil{National Radio Astronomy Observatory, Charlottesville,
 VA, (bcotton@nrao.edu)}

\begin{abstract}

The absolute position of the compact radio source at the dynamical center 
of the Galaxy, Sgr A$^*$,  was known only to an accuracy of 
$0.2''$
in spite of its accurate location with respect to  near-IR stellar 
sources to within 30 milliarcsecond (mas). 
To remedy this poor positional accuracy,  we have  selected 15
high-resolution, high-frequency  VLA  observations 
of Sgr A$^*$  carried out in  the last 13 years and determined 
the weighted average position with the average epoch 1992.4 to be at 
$\alpha$, $\delta$[1950] = $17^{\rm h} 42^{\rm m}$ 29\dsec3076$\pm0.0007$,
$-28^\circ 59^\prime 18.484\pm0.014^{\prime\prime}$, or
$\alpha$, $\delta $[2000] = $17^{\rm h} 45^{\rm m}$
40\dsec0383$\pm0.0007$, $-29^\circ 00^\prime 28.069\pm0.014^{\prime\prime}$
which agrees with earlier published values to within the $0.2''$ error
bars of the earlier measurements.  
An  accurate absolute position of Sgr A$^*$ can be useful
for its identification with  sources at other wavelengths,
particularly,  in soft and hard X-rays
with implications for the models  of a massive black hole at the
Galactic center.  

\end{abstract}


\vfill\eject

\section{Introduction}

Recent advances in infrared and X-ray array  technologies  have greatly increased
the spatial coverage and resolutions in near-IR and X-ray
observations, providing powerful tools to compare high-resolution 
radio,  IR  and X-ray images of the complex region of the Galactic
center. Sgr A$^*$ is a bright compact radio source
which  is considered to be the strongest candidate 
for  a 2.5$\times10^6$ \msol black hole at the 
dynamical center of the Galaxy (Eckart and Genzel 1997; Ghez et al. 1998). 
This source is surrounded by diffuse ionized gas as well as  clusters 
of  hot and cool stars seen in near-IR wavelengths. 
Detailed comparison of radio and near-IR images 
was recently made with a new approach by 
Menten et al. (1997), who detected a number of SiO (0.7 cm) and H$_2$O
(1.3 cm) masers associated with known compact stellar sources seen in
the diffraction limited 2.2 $\mu$m infrared images.  
Astrometric measurements between radio and 2 $\mu$m images of this region 
were  used to determine the position of Sgr A$^*$ with respect to other 
well-known near-IR sources. 
Although the   relative position of Sgr A$^*$ 
is accurately measured within 30 mas in this technique, the absolute position of
Sgr A$^*$ was  known  with an   accuracy of 0.2$''$. 
A more accurate  position of this  source and its subsequent X-ray counterpart 
in future sub-arcsecond observations has important implications on the nature of 
high energy activity predicted from models of this source (Melia 1994;
Narayan et al. 1995).

The absolute positions of radio sources are routinely obtained to an
accuracy of a few milliarcseconds using centimeter wavelength VLBI
techniques. 
However, the large scattering diameter of Sgr A$^*$ (Yusef-Zadeh et
al. 1994) makes the application of centimeter wavelength VLBI less
feasible, as the source is completely resolved on longer baselines.
Thus, shorter baseline arrays and/or shorter wavelengths are needed
resulting in reduced positional accuracy.

The most recent measurements of the location of Sgr A$^*$ with the
smallest error bar is given as $\alpha$, $\delta$[ 1950 ] = $17^{\rm
h} 42^{\rm m}$ 29\dsec314$\pm0.010$, 
$-28^\circ 59^\prime 18.3\pm0.2^{\prime\prime}$ 
(Rogers et al. 1994). 
This position is determined with respect to the reference source NRAO 530 
using VLBI techniques at 3mm.

To improve the absolute position of Sgr A$^*$, we obtained archival
VLA data sets at a number of frequencies taken over the last 13 years.
By averaging the position of Sgr A$^*$ at different frequencies and 
at different epochs, the new 
position of  Sgr A$^*$ is determined with an accuracy of an order of
magnitude better than earlier measurements. 
The averaging of the data over such a long span of time and frequency
can be justified.  
The proper motion measurements of this source is too small  with
respect to reference extragalactic radio sources (Reid et al. 1998;
Backer 1996) to affect individual  measurement.  
Also, the change in the position of Sgr A$^*$ as a function of frequency is 
too small to be significant compared to other systematic errors.
The frequency dependence of the size of Sgr A$^*$ due
to scattering of radio waves will not change the centroid of the 
position of Sgr A$^*$ as a function of frequency (Yusef-Zadeh et al. 1994). 
It is worth mentioning 
that the detection of frequency dependence of the position of Sgr A$^*$ 
at a milliarcsecond level, due to its intrinsic radio emission, 
has the potential to support the asymmetrical Bondi-Hoyle accretion model
(Melia 1994). 

\section{Observations}

Table 1 shows the list of 15 data sets that were based on 
continuum observations
made with the  Very Large Array of the National
Radio Astronomy Observatory\footnote{The National Radio Astronomy
Observatory is a facility of the National Science Foundation, operated
under a cooperative agreement by Associated Universities, Inc.} at 
6 (5 GHz), 3.6 (8.4 GHz), 2 (15 GHz), 1.2 (20 GHz), and 0.7 cm (44
GHz) between 1986 and 1999. 
All had a bandwidth ranging between  2 $\times$ 50 and  2 $\times$ 12.5 
MHz in each right and left
circular polarization. 
Each of these original data sets  had relatively good  {uv} coverage
and none were done in snapshot mode with less than one hour of
integration time on the source. 
With the exception of 0.7 cm data which was taken in the C configuration,
 all the observations were done in either
the A or the B
configurations of the VLA to obtain the highest resolution images of
Sgr A$^*$ with minimal  contamination or confusion by ionized features
in the vicinity of Sgr A$^*$.
The low resolution images of Sgr A$^*$ at 20 and 90 cm suffer from
this source confusion as well as optical depth effects due to
intervening thermal and nonthermal continuum sources. 
Thus, 20 and 90cm  observations were not used. 

In all observations, NRAO530 and 3C286 were  chosen as the  phase
(astrometric) and flux (photometric) calibrators, respectively.  
All processing was done using the NRAO AIPS analysis package; the
standard calibration was applied and the field around Sgr A$^*$ was
imaged.
Since the observations were all made in the B1950 coordinate system,
the data were analyzed in this system.  The VLA calibrator B1950
coordinate system is referred to epoch 1979.5.
No self--calibration was done on any of the data as this procedure
loses the absolute position of a source in order to minimize
atmospheric phase errors. 
Prior to self-calibration, no significant emission from Sgr A$^*$ was
detected at 0.7 cm in the A and B configurations, so these data were
excluded.  
VLA continuum data consists of two independent frequency bands.
The final images based on combining these two frequency bands were
used to determine the position of Sgr A$^*$ with AIPS routine JMFIT
by fitting a Gaussian to the source.
Fits done using images from the individual frequency bands showed no
significant positional shift.  

At short wavelengths, the phase fluctuations degrading the positional
accuracy are due mainly to variable refraction in the troposphere; at
longer wavelengths the ionosphere adds increasingly important
contributions.
Variable ionospheric refraction is particularly significant for
observations made at 6 cm during the day when the ionosphere is
densest.
Solar heating of the troposphere also increases the variations in
refraction.
So, we have assigned weights of 1 and 2  to day-- and night-time
observations, respectively.
For observations that were carried out between night and day, the
assigned  weight was 1.5. 

The atmospheric effects are aggravated by the low declination of the source 
and the
calibrator and the resultant low elevation angles of the observations.
These effects are  minimized by restricting  the {\it uv} data 
to near transit. 
The data  sets selected within one to three hours
 of the highest elevation
improved considerably the dispersion in  measured position of Sgr
A$^*$ among the different epochs. 

Errors in atmospheric refraction  affect mainly the  measurements of
the declination of Sgr A$^*$.
The astrometric calibrator used, NRAO530, is about 15$^\circ$ north of
Sgr A$^*$.
Since the observations were restricted to those near transit,
this corresponds to a difference primarily in elevation.
Corrections to the atmospheric refraction model are determined from
the calibrator observations which is at a significantly higher
elevation than Sgr A$^*$ (i.e. at lower air mass); any unmodeled
differences in atmospheric refraction between the calibrator and Sgr
A$^*$ will appear primarily as errors in declination.

A further potential source of systematic error is the assumed position of
the calibrator, NRAO530.  Our measurements are of position relative to
that of the calibrator whose absolute position is known; any errors in the
assumed position of NRAO530 will directly affect the derived position of
Sgr A$^*$.  In all our measurements, the position of NRAO530 was constant
and was assumed to be at (B1950 epoch 1979.5) $\alpha$, $\delta$ =
$17^{\rm h} 30^{\rm m} 13\dsec5352$, $-13^\circ 02^\prime
45.837^{\prime\prime}$. This position is consistent to within 5 mas of the
VLBI source position given by Johnston et al. (1995) and by 
US Naval Observatory's (USNO) astrometry
list (ftp://casa.usno.navy.mil/navnet/n9798.sor) after the correction
between the VLA and USNO frames is made.  

The position of NRAO530 assumed is an absolute position measured by
VLBI techniques which are insensitive to structure larger than a few
tens of milliarcseconds.
NRAO530 is suspected of having structure on the scale of several
hundred milliarcseconds which would not affect the VLBI measurements
but will cause an apparent shift of the position when viewed with the
lower VLA resolution.
Any difference between the centroid of the position of NRAO530 at
VLBI and VLA resolutions will appear as an error in the derived
position of Sgr A$^*$.

\section{Results}

Table 1 lists 15 data sets, three of which at 1.3 cm, four at 
2 cm, five at 3.6 cm, two  at 6 cm, and one at 0.7 cm.
The date of each observation,  the Gaussian fitted position with 1
$\sigma$ error, duration of the observation
 and the corresponding weight in the final averaging
are also given. 
The errors of the fitted positions depend on the signal-to-noise
ratio and are much  less than the errors introduced 
by the atmosphere. The rms scatter within the data set, as described
below, is about ten times the fitted noise, thus we have not 
used the errors of the fitted positions in any of our 
analysis. 
The mean  position of Sgr A$^*$, both weighted and unweighted, 
as well as the position determined by Rogers et al. 1994 are also
listed  at the bottom of the table.  The most descripant point 
at 1.3 cm was not included in the final averaging. 

   The rms  scatter of the individual measurements in Table 1
(either weighted or
unweighted) is about 0.035$''$ in Right ascension and 0.055" in
declination.  This is consistent with the assumption that the errors
are dominated by uncertainties in atmospheric refraction which should
produce a larger scatter in declination.  This is further supported by
the two pairs of positions measured on the same day, in which the
declinations differed by about 0.030", or nearly half of the day-to-day
scatter.  The one sigma errors in the mean position are taken to be the
rms  scatter of the sample divided by the square root of the number of
observations minus one.

Figure 1 shows the Right ascension and declination  of Sgr A$^*$ with
one sigma error given  by Rogers et al. (1994) against 
15 positions determined from our measurements. 
All the new positions fall   to the SW quadrant of the
position derived by  Rogers et al. 
The weighted position of Sgr A$^*$ with the weighted 
mean  epoch of  our observation being 1992.4 is \hfill\break
(B1950) $\alpha$, $\delta$ =
$17^{\rm h} 42^{\rm m}$ 29\dsec3076$\pm0.0007$, $-28^\circ 59^\prime
18.484\pm0.014^{\prime\prime}$ or \hfill\break
(J2000) $\alpha$, $\delta$ =
$17^{\rm h} 45^{\rm m}$ 40\dsec0383$\pm0.0007$, $-29^\circ 00^\prime
28.069\pm0.014^{\prime\prime}$. \hfill\break
These measurements are in  good agreement with the less precise measurement
of Rogers et al. 1994 as well as with the 
position of Sgr A$^*$ determined from recent VLBI measurements (Reid et
al. 1999).

\acknowledgments {We thank D. Roberts for his help in  calibrating
0.7 cm data and the referee, Mark Reid, 
 for sharing with us the VLBI  position of Sgr
A$^*$ prior  to publications and for his useful comments. }
 \vfill\eject

\begin{figure}
\plotone{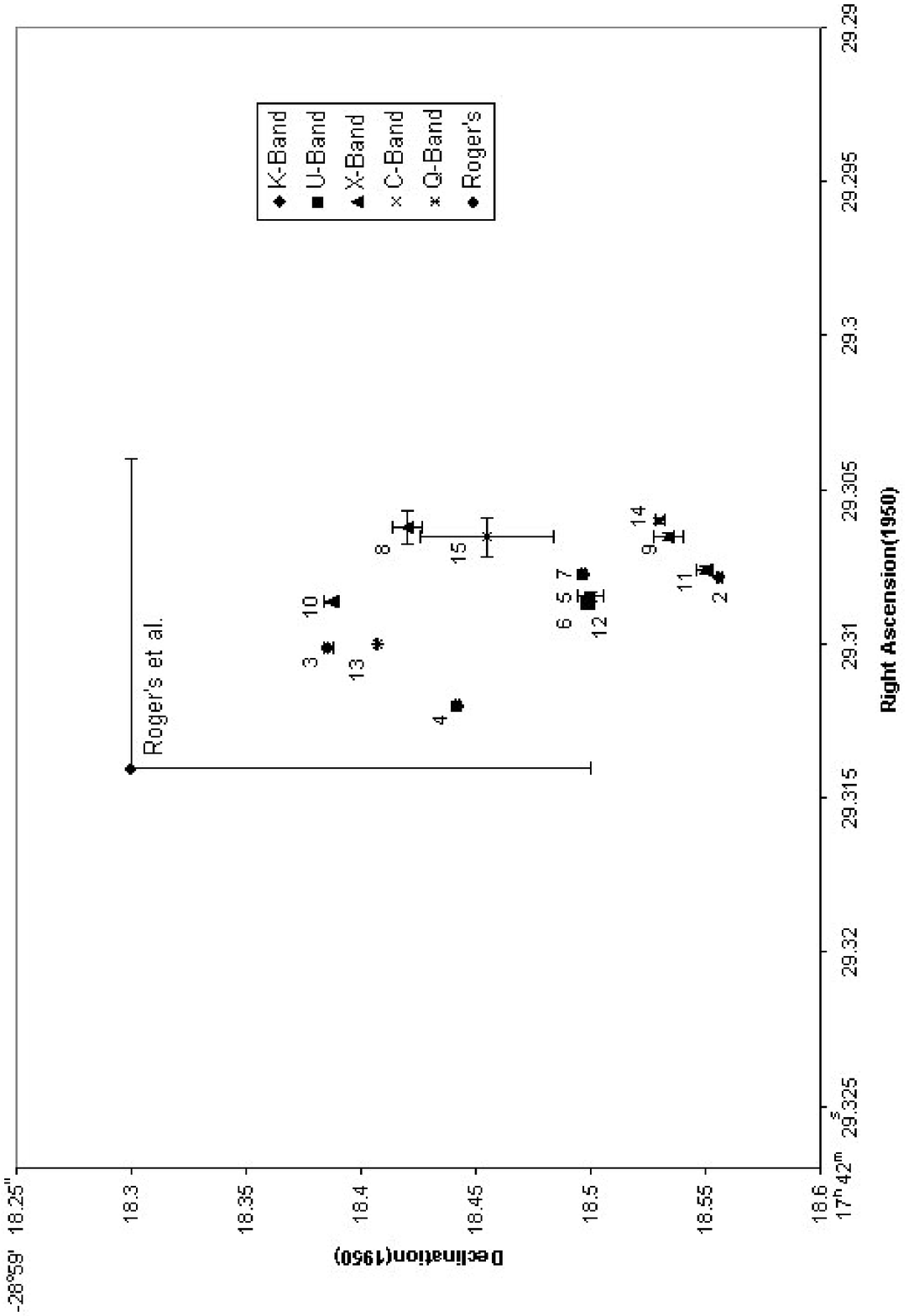}
\figcaption{ The VLBI position of Sgr A$^*$ and its error bar, 
as given by Rogers et al. (1994),   is  plotted against VLA positions
of Sgr A$^*$ and their uncertainties based on 15 different measurements. The 
numbers correspond to data set designations shown in Table 1.} 
\end{figure}

\begin{table}
\plotone{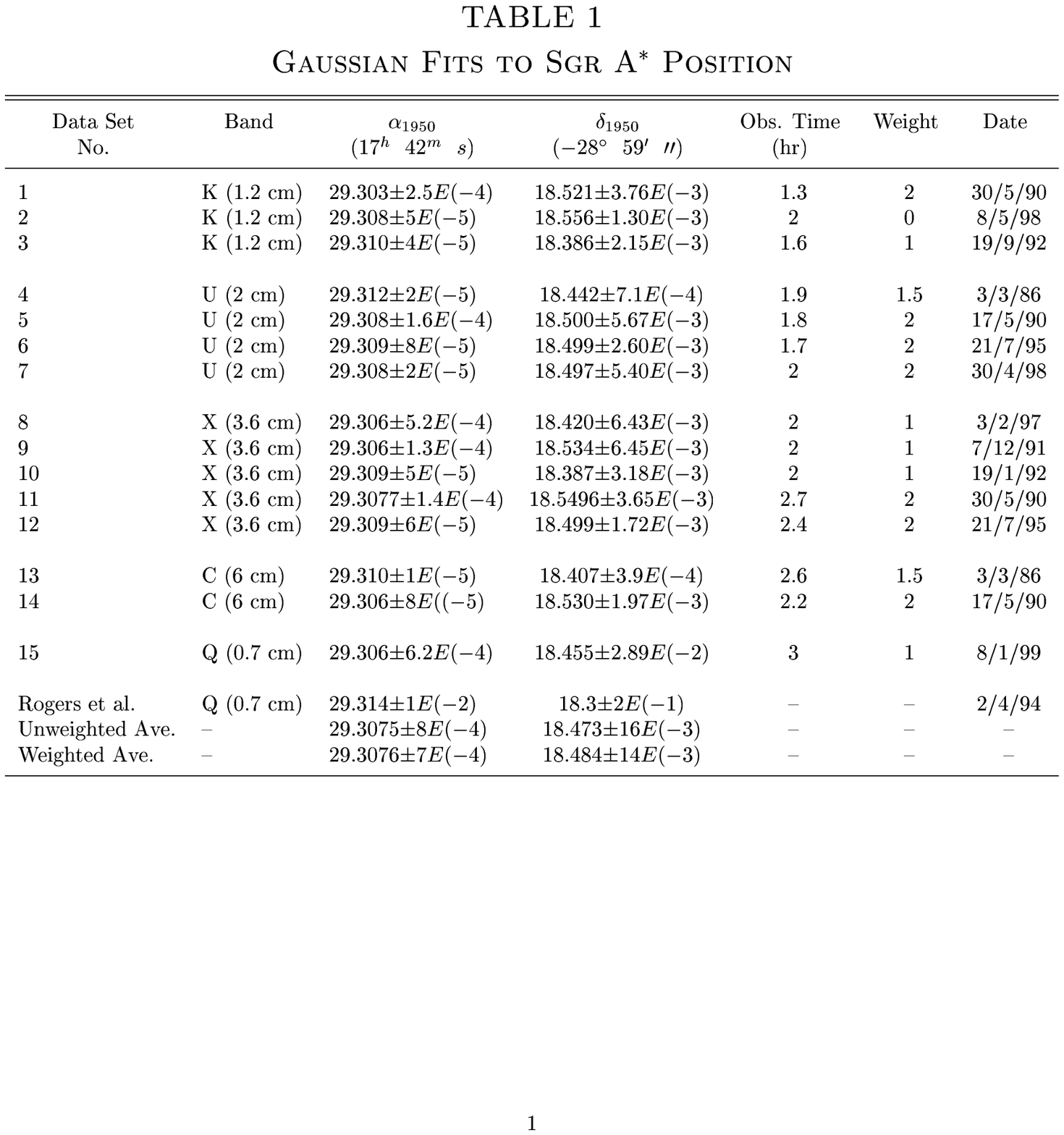}
\figcaption{ Table 1}
\end{table}

\end{document}